\documentclass[conference]{IEEEtran}
\IEEEoverridecommandlockouts
\usepackage{cite}
\usepackage{amsmath,amssymb,amsfonts}
\usepackage{algorithmic}
\usepackage{graphicx}
\usepackage{textcomp}
\usepackage{xcolor}
\usepackage{float}
\usepackage[belowskip=-3pt]{caption}
\def\BibTeX{{\rm B\kern-.05em{\sc i\kern-.025em b}\kern-.08em
    T\kern-.1667em\lower.7ex\hbox{E}\kern-.125emX}}
\newcommand\Mark[1]{\textsuperscript#1}
\begin{document}

\title{Adaptation and Attention for Neural Video Coding}

\author{Nannan Zou\Mark{{1, 2}}, Honglei Zhang\Mark{1}, Francesco Cricri\Mark{1}, Ramin G. Youvalari\Mark{1}, Hamed R. Tavakoli\Mark{1} \\
Jani Lainema\Mark{1}, Emre Aksu\Mark{1}, Miska Hannuksela\Mark{1}, Esa Rahtu\Mark{2}\\
\Mark{1}Nokia Technologies, \Mark{2}Tampere University, Tampere, Finland\\
{\tt\small nannan.zou@nokia.com}}

\maketitle

\begin{abstract}
Neural image coding represents now the state-of-the-art image compression approach. 
However, a lot of work is still to be done in the video domain. In this work, we propose an end-to-end learned video codec 
that introduces several architectural novelties as well as training novelties, revolving around the concepts of adaptation and attention. 
Our codec is organized as an intra-frame codec paired with an inter-frame codec. As one architectural novelty, 
we propose to train the inter-frame codec model to adapt the motion estimation process based on the resolution of the input video. 
A second architectural novelty is a new neural block that combines concepts from split-attention based neural networks and from DenseNets. 
Finally, we propose to overfit a set of decoder-side multiplicative parameters at inference time. Through ablation studies and comparisons to prior art, we show the benefits of our proposed techniques in terms of coding gains. 
We compare our codec to VVC/H.266 and RLVC, which represent the state-of-the-art traditional and end-to-end learned codecs, respectively, 
and to the top performing end-to-end learned approach in 2021 CLIC competition, 
\texttt{E2E\_T\_OL}. Our codec clearly outperforms \texttt{E2E\_T\_OL}, and compare favorably to VVC and RLVC in some settings. 

\end{abstract}


\section{Introduction}
Nowadays, image codecs based on deep learning represent the state-of-the-art when considering MS-SSIM \cite{Cheng2020} 
and PSNR \cite{Guo2021} visual quality metrics. The typical architecture is based on the auto-encoder, 
where the encoder and decoder neural networks perform non-linear forward and inverse transform, respectively. 
The output of the encoder is typically lossless encoded by an arithmetic encoder, using a learned probability model. 
In the video domain, however, the state-of-the-art is represented by traditional codecs such as VVC/H.266 \cite{vvc2021} 
and HEVC/H.265 \cite{Sullivan2012a} standards, which follow the prediction-transform paradigm: intra-frame and inter-frame prediction, 
followed by transform-coding of prediction residuals. We propose a learned codec that follows a common design \cite{Wu2018,Yang2020} 
inspired by traditional codecs, where a learned image codec performs intra-frame coding, 
and a conditional interpolation model interpolates the other frames based on reconstructed intra frames. 
Based on the observation that the extent by which objects move in terms of pixels depends also on the spatial resolution, 
in our inter-frame codec, the output of motion estimation is adapted by the input video's resolution.
Recently, efficient attention blocks have been proposed, such as the Split Attention block (or ResNeSt block) \cite{Zhang2020}, 
which applies the squeeze-and-attention concept \cite{Hu2018} \cite{Li2019} to groups of feature maps. 
ResNeSt blocks have already been successfully used in \cite{Guo2021} as one of the blocks within a learned image codec. 
We further extend the ResNeSt block idea by designing the Dense Split Attention (DSA) block, 
that combines the efficiency of split attention with the power of dense connections \cite{Huang2017} 
between each of a set of ResNet blocks and the output of split attention.
To further optimize part of the codec to the input content at inference time, 
several prior works have proposed to optimize or \textit{overfit} 
some of the encoder's parameters \cite{Aytekin2018a} \cite{Aytekin2019} \cite{Lu2020}, 
or the latent tensor output by the encoder \cite{Campos2019} \cite{Zou2020}, or some or 
all the decoder's parameters \cite{HongLam2019} \cite{Lam2020}, or both latent tensor 
and decoder's parameters \cite{vanrozendaal2021overfitting}. 
Overfitting decoder's parameters would incur into a bitrate overhead for providing the weight-update to the decoder side. 
To limit such overhead in the case of a VVC codec augmented with a decoder-side post-processing neural network, 
the authors of \cite{Lam2020} propose to overfit only the bias terms of all the convolutional layers 
on all the frames in a Group of Pictures (GOP). Our overfitting process is applied on multiplicative parameters 
instead of additive bias terms. Also, we argue that \textit{(i)} the layers of a neural network are not equally important, 
thus we overfit only a selected subset of parameters, and \textit{(ii)} overfitting on a single frame is enough for short sequences without scene changes.

In summary, we propose a highly-adaptive neural video codec, which includes the following contributions: 
\textit{(i)} resolution-adaptive motion estimation, \textit{(ii)} Dense Split Attention blocks, 
\textit{(iii)} overfitting of multiplicative parameters on a single frame per video, 
\textit{(iv)} adapting the combination of forward and backward predictions on each frame.

\section{Proposed Methods}
Our video codec is organized as an intra-frame codec and an inter-frame codec. 
The intra-frame codec processes one frame every eight frames, without using any information from other frames. 
The seven frames between two consecutive intra-frames are coded by the inter-frame codec in a hierarchical sequential manner. 
I.e., first the frame with index $4$ (relative to the start of the intra-frame period) is coded based on intra-coded frames $\{0,8\}$, 
then frame $2$ and frame $6$ are coded based on intra or inter-coded frames $\{0,4\}$ and $\{4,8\}$, respectively, 
finally frames $1$, $3$, $5$ and $7$ are coded based on intra or inter-coded frames $\{0,2\}$, $\{2,4\}$, $\{4,6\}$ and $\{6,8\}$, respectively.

\subsection{Intra-Frame Codec}
\begin{figure*}
\begin{center}
\includegraphics[width=14cm, height=4.7cm]{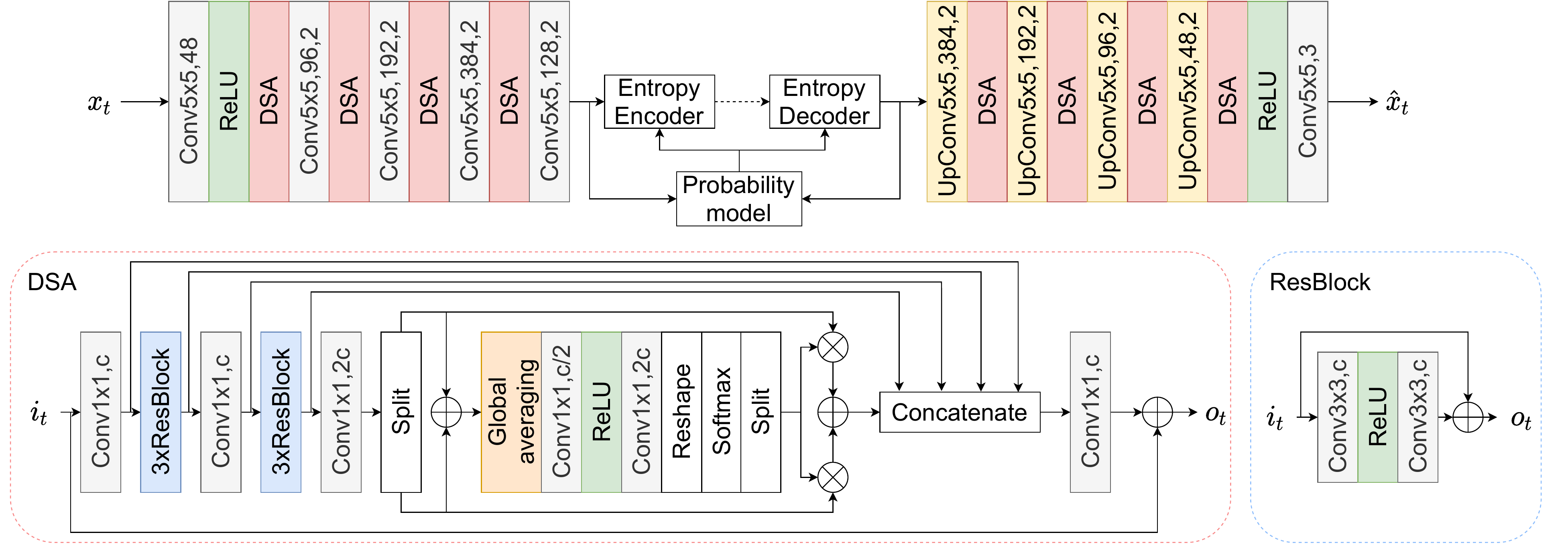}
\end{center}
   \caption{Our intra-frame codec, with the proposed Dense Split Attention (DSA) block. "Conv$K{\times}K$,$c$,$s$" and "UpConv$K{\times}K$,$c$,$s$" stand for 2D convolutional layer and 2D transposed convolutional layer, respectively, with kernel size $K{\times}K$, $c$ output channels, and an optional stride value $s$ if it is not 1.}
\label{fig:intra_codec}
\end{figure*}

Our intra-frame codec, shown in Figure \ref{fig:intra_codec}, has an autoencoder architecture similar to other end-to-end 
learned image codecs \cite{Cheng2020,Liu_Chen_Guo_Shen_Cao_Wang_Ma_2019,Minnen_Balle_Toderici_2018}. 
The encoder transforms the input frame to a latent representation, which is quantized using uniform scalar quantization 
(the quantizer is not shown for simplicity). The Probability Model component estimates the probability 
of each element in the latent representation, and provides it to the entropy codec, which is an arithmetic codec. 
The decoder has a mirrored architecture with respect to the encoder.

One novel component of our intra-frame codec is the Dense Split Attention (DSA) block, 
which is shown in detail in Figure \ref{fig:intra_codec}. This block combines the efficiency of 
split channel attention \cite{Zhang2020} with the power of dense connections \cite{Huang2017}. 
In particular, DSA consists of an initial set of convolutional layers and ResBlock layers, 
followed by a core attention block whose output is concatenated with dense connections coming from the initial set of layers, 
before being processed by a final convolutional layer. The core attention block can be any attention-based block, 
such as non-local attention \cite{Liu_Chen_Guo_Shen_Cao_Wang_Ma_2019}. 
However, in order to keep computational and memory consumption low, 
we opted for the efficient Split Attention block \cite{Zhang2020} with $k=1$ group and $r=2$ splits of features.

Another novel aspect of our intra-frame codec is the Overfittable Multiplicative Parameters (OMPs). 
An OMP is a learnable parameter that multiplies a feature map output by a certain kernel of a convolutional layer. 
The OMPs are initialized to $1$ and then optimized at inference time. 
However, we chose to use OMPs only within the last DSA block of the intra-decoder, 
more specifically on the output of the following layers: the first convolutional layer, 
the first convolutional layer of the first ResBlock, the second convolutional layer of the second and fourth ResBlock. 
These layers were selected empirically based on an evaluation on validation data. 

\subsection{Inter-Frame Codec}
\begin{figure*}
\begin{center}

\includegraphics[width=17cm]{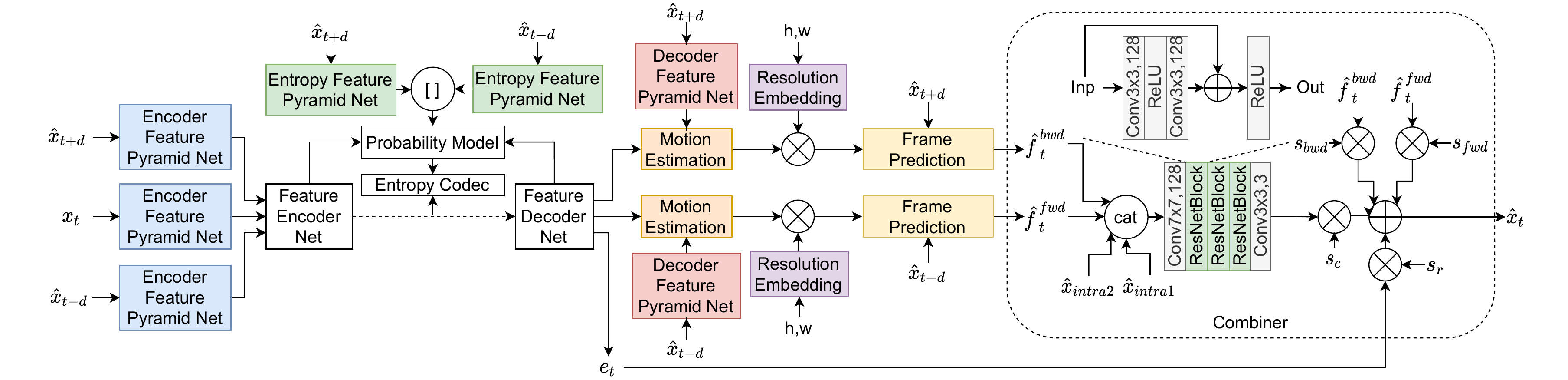}
\end{center}
   \caption{Architecture of the inter-frame codec. In this figure, components that share weights are shown with the same color.}
\label{fig:inter_codec}
\end{figure*}

An overview of the inter-frame codec is provided in Figure \ref{fig:inter_codec}. 
The inputs to the encoder are two reconstructed reference frames $\hat{x}_{t-d}$, $\hat{x}_{t+d}$ and the target frame $x_t$.
In practice, we define $d$ as the distance from the target frame $x_t$, which is randomly chosen among \{1, 2, 4\}.
A reconstructed reference frame may be an intra-coded frame or a inter-coded frame. 
First, the Encoder Feature Pyramid Net extracts multi-scale features from the input frames. 
Next, the multi-scale features are aggregated by the Feature Encoder Net and transformed to a latent representation of the target frame. 
Then, the Entropy Encoder, in this case an arithmetic encoder, 
compresses the latent representation into a bitstream by using the output of the Probability Model. 
Our probability model is conditioned on features extracted from the reconstructed reference frames, by using the Entropy Feature Pyramid Net. 
The decoder takes the compressed bitstream of the latent representation as its input. 
The bitstream is first decompressed and dequantized by the Entropy Decoder to generate a reconstruction of the latent representation. 
The latent representation implicitly embeds forward motion information (from $\hat{x}_{t-d}$ to $\hat{x}_{t}$), 
backward motion information (from $\hat{x}_{t+d}$ to $\hat{x}_{t}$), and information about a residual signal. 
The reconstructed latent tensor is passed to the Feature Decoder Net that generates multi-scale motion features and a residual signal $e_t$. 
The Decoder Feature Pyramid Net extracts multi-scale features from reference frames. 
The Motion Estimation module takes in the multi-scale motion features and reference frame's multi-scale features to output motion information. 
The Motion Estimation has a similar architecture as FlowNet \cite{flownet_2015,Yang2020}. 
Different from other systems \cite{Liu_Lu_Ma_Wang_Xie_Cao_Wang_2020}, 
the Motion Estimation is randomly initialized and end-to-end learned together with other components in our system. 
The resolution information, represented by the height $h$ and width $w$ of the video frames, 
is embedded into feature space by the Resolution Embedding module. The embedded resolution is used to scale the motion information. 
The scaled motion information is then used by the Frame Prediction module to warp the reference frames into a forward prediction $\hat{f}_{t}^{fwd}$ 
and a backward prediction $\hat{f}_{t}^{bwd}$ of the target frame. 
These predictions are then combined with the residual and with the two closest intra-coded frames 
by the Combiner component to produce the final reconstructed target frame $\hat{x}_t$. 
We leave out details about several of the above modules because of space limitations. 
They are based on known neural network architectures and do not include major novelties. 
The Resolution Embedding module consists of two fully-connected layers followed by a Leaky ReLU layer. 
Regarding the Combiner, the input forward and backward predictions of the target frame $\hat{f}^{fwd}_t$ and $\hat{f}^{bwd}_t$ 
are concatenated with the closest intra-coded frames $\hat{x}_{intra1}$ and $\hat{x}_{intra2}$, 
and then processed by a sequence of convolutional layers and ResNet blocks \cite{He_Zhang_Ren_Sun_2015}, 
obtaining a temporary prediction $\tilde{x}_t$. Then, $\tilde{x}_t$, $\hat{f}^{fwd}_t$, $\hat{f}^{bwd}_t$ 
and the decoded residual $e_t$ are combined by a linear combination
 $\hat{x}_{t}=s^{tmp}_t \tilde{x}_{t} + s^{bwd}_t \hat{f}^{bwd}_t + s^{fwd}_t \hat{f}^{fwd}_t + s^e_t e_t$, 
 where $s^{tmp}_t$, $s^{bwd}_t$, $s^{fwd}_t$, $s^e_t$ are scalars that are trained during the training stage of the inter-frame codec, 
 and overfitted at inference time.


\subsection{Probability Model}
Our probability models used in the intra-frame and inter-frame codec are based on the state-of-the-art 
probability model for lossless image compression described in \cite{zhang2020lossless}. 
An input latent representation is first downscaled to multiple resolutions. 
The representation in the lowest resolution is compressed/decompressed using a non-conditional distribution 
model with the assumption that the elements are independent and identically distributed. 
The distributions of the representations in other resolutions are conditioned on the representation in lower resolutions. 
At the encoding and decoding stage, the system first processes the representation in the lowest resolution. 
Then, the system moves to the next higher-resolution representation, 
using the low-resolution representation as a context to derive the distribution function. 
This procedure continues until the original latent representation is processed. 
To further improve the performance, we partition the elements in each representation into several groups. 
The elements in one group are processed together using the elements that have been processed as a context. 
We adopt the conditional Gaussian distribution model as used in \cite{meanscale_hyperprior} 
for the sake of accuracy and computational efficiency. The parameters of the distribution model, 
i.e., means and scales, are estimated using a deep neural network given the input of the low-resolution representation 
and the elements in the same resolution that have been processed. The probability model used in the inter-frame codec 
is enhanced by taking features extracted from the reconstructed reference frames as extra context for representations 
in resolutions other than the lowest one.

\subsection{Training and Inference Aspects}
Both the intra-frame codec and the inter-frame codec were trained on 256x256 patches, 
by using MS-SSIM and rate loss as the training objectives: $\mathcal{L} = \mathcal{D} + \lambda \mathcal{R}$, 
where the distortion $\mathcal{D}$ is the negative MS-SSIM, $\mathcal{R}$ is the rate derived from the probability model, 
and $\lambda$ is a hyper-parameter. The intra-frame codec was trained only as a stand-alone module. 
The inter-frame codec was first pretrained as a stand-alone module, by using uncompressed reference frames, 
and then finetuned by using a similar pipeline as at inference time, i.e., by using intra-coded frames and 
inter-coded frames as reference frames in hierarchical sequential processing. 

At inference time and for each intra frame, the latent tensor output by the intra-frame encoder is overfitted 
as in \cite{Zou2020}. After that, the OMPs of the intra-frame decoder are optimized on the first intra frame of each video, 
and used for all intra frames of that video. This strategy leverages the high temporal redundancy of videos when there is no scene change. 
A strategy that overfits the OMPs 
jointly on all intra frames of a video would be more time-consuming, and the coding gains may not be worth the extra time. 
Another alternative strategy would overfit a separate set of OMPs for each intra frame. However, coding gains would be 
negatively affected by a much higher bitrate overhead required by the overfitted parameters. 
In the experimental section, we provide comparisons for some of these strategies. Based on our comparisons, we 
choose the first optimization strategy mentioned above for our intra codec decoder. 
After overfitting, the overfitted OMPs are uniformly quantized to 8 bits. If the quantized OMPs provide coding gains 
in terms of the overall loss over a video, the parameters would work as separate bitstreams together with latent tensor bitstreams  
for decoding. Otherwise, we do not include them into the bitstreams.  
We also propose to overfit, at inference time, the scaling parameters used within the Combiner module of the inter-frame decoder. 
These parameters are adapted separately on each inter-coded frame. 


\section{Experiments}
The codec was trained on the CLIC 2021 video dataset. 
The intra-frame codec was trained by using all the frames of all the training videos for $60$ epochs, 
with a learning rate of $5e{-}5$ and batch-size of $60$ frames. 
The inter-frame codec was first pretrained on uncompressed reference frames 
for which the distance from the target frame was randomly chosen among $\{1,2,4\}$. 
This pretraining was performed for $34$ epochs, a learning rate of $5e{-}5$ and 
a batch-size of $63$ samples, where each sample consists of two reference frames and one target frame. 
The inter-frame codec is then finetuned on all frames of all videos in the dataset, 
for $10$ epochs, with a learning rate of $2e{-}5$ and batch-size of $56$ sets of $7$ inter frames.
We follow the evaluation framework of the CVPR workshop and challenge CLIC (video track), which is the most recent learned video coding conference, to allow for an easier comparison with prior art learned codecs. According to this framework, the combined size of the decoder and bitstreams, calculated as
$decodersize + bitstreams / 0.019$, should not exceed $1309$MB. 
We tested our codec on the CLIC test set and on the JVET-CTC sequences. 
For JVET-CTC sequences, we excluded Class A due to the high resolution causing high memory consumption, 
and we converted 10 bits sequences to 8 bits for simplicity. 
We compared our codec to the state-of-the-art traditional and learned video codec, 
i.e., VVC/H.266 and RLVC \cite{yang2021learning}, respectively.

For VVC, the VTM-12.0 software was used in our comparison. 
We tuned its hyper-parameters to achieve the target combined size on the CLIC dataset as close as possible. 
We evaluated it with an intra period of 8 frames (same as our codec) on CLIC test set, 
and both an intra period of 8 frames and an intra period of 1 second (default setting) on JVET-CTC sequences. 
To evaluate RLVC, the bi-IPPP GOP structure with default settings was adopted in our experiments. 
Additionally, we used their MS-SSIM model with lambda value of 8 which provides the smallest bitrate. 
Nonetheless, RLVC still cannot achieve the target combined size, as showed in Table \ref{tab:results}.
As RLVC was trained on RGB data, and the test datasets are in YUV 4:2:0 color format, we converted the videos to RGB by using FFmpeg. 
We measured the quality drop caused by the conversion as the MS-SSIM and the peak signal-to-noise ratio (PSNR) 
computed on the original YUV data and corresponding YUV data obtained after converting to RGB and back to YUV. 
For CLIC test set, the MS-SSIM was $0.994$, and the PSNR was $55.9$ dB. 
For JVET-CTC sequences, the MS-SSIM was $0.999$, and the PSNR was $50.5$ dB. 
We provide test results both in RGB and YUV domains. 

Table \ref{tab:results} reports the results for the above experiments. 
NNVC refers to the proposed codec. VVC8 is VVC with intra period of 8 frames. 
RLVC-RGB and RLVC-YUV are RLVC evaluated in RGB and YUV domain, respectively.
We also report the performance of the top performing end-to-end learned approach in 2021 CLIC competition, \texttt{E2E\_T\_OL}. 
NNVC surpasses their performance.
We measure the speed of NNVC on one NVIDIA Tesla V100 SXM2 GPU. 
For an 1280x720 CLIC video, our encoding (including the overfitting) and decoding run on average at 0.006 and 1.2 frames/sec, respectively.
For the same video, VVC encodes and decodes on average at 0.022 and 23.6 frames/sec, respectively, with one Intel Xeon Gold 6154 CPU.

\begin{table}
\caption{Experimental results. 
}
\begin{center}
\begin{tabular}{|l|c|c|c|c|}
\hline
\textbf{Model} & \textbf{Data} & \textbf{BPP} & \textbf{MS-SSIM} & \textbf{Combined Size}\\
\hline\hline
NNVC & CLIC & $0.03095$ & $0.97347$ & $1306$MB \\
VVC8 & CLIC & $0.03487$ & $0.97105$ & $1298$MB \\
RLVC-RGB & CLIC & $0.06752$ & $0.97056$ & $2663$MB \\
RLVC-YUV & CLIC & $0.06752$ & $0.97494$ & $2663$MB \\
E2E\_T\_OL & CLIC & $0.03395$ & $0.97167$ & $1306$MB \\
\hline
NNVC & JVET & $0.03772$ & $0.97001$ & $2197$MB \\
VVC8 & JVET & $0.03693$ & $0.97461$ & $1999$MB\\
VVC & JVET & $0.03788$ & $0.98576$ & $2051$MB\\
RLVC-RGB & JVET & $0.07332$ & $0.96927$ & $4120$MB \\
RLVC-YUV & JVET & $0.07332$ & $0.98178$ & $4120$MB \\
\hline
\end{tabular}
\label{tab:results}
\end{center}
\end{table}

Regarding the intra-frame codec, we compare our DSA block with the "group-separated attention (GSA)" 
block used in the prior art learned image codec \cite{Guo2021}. 
To this end, we designed a codec which is as similar as possible to our proposed codec, but which uses the GSA block. 
While DSA's ResBlock uses two convolutional layers, 
in \cite{Guo2021} the residual block uses three convolutional layers. 
In our study, we compared both of these two cases.
Table \ref{tab:DSA_results} describes the results of comparison. 
The "Score" values are derived as the negative of the loss values computed on the evaluation set by using $\lambda=0.1$. 
A higher score value indicates better performance. 
DSA-N and GSA-N refer to the proposed NNVC's intra codec and the NNVC's intra codec 
where DSA blocks were replaced with GSA blocks \cite{Guo2021}, respectively. 
N is the number of convolutional layers in the ResBlock of DSA or GSA. 
The models were trained on 30 videos of the CLIC training set, 
and evaluated on $10\%$ of the frames in those videos (similar setup as in CLIC competition).
The results clearly show that DSA block performs better than GSA, even when GSA uses more layers and parameters (i.e., DSA-2 vs GSA-3).

\begin{table}
\caption{Comparison between DSA and GSA blocks. 
} 
\begin{center}
\begin{tabular}{|l|c|c|c|c|c|}
\hline
\textbf{Model} & \textbf{BPP} & \textbf{MS-SSIM} & \textbf{Score} & \textbf{Parameters}\\
\hline\hline
DSA-2 & $0.13168$ & $0.98857$ & $0.97540$ & $55$M \\
DSA-3 & $0.13182$ & $0.98883$ & $0.97565$ & $76$M \\
GSA-2 & $0.13355$ & $0.98837$ & $0.97502$ & $52$M \\
GSA-3 & $0.13046$ & $0.98843$ & $0.97538$ & $73$M \\
\hline

\end{tabular}
\label{tab:DSA_results}
\end{center}
\end{table}

In another experiment (see \textbf{Intra codec only} part in Table \ref{tab:study_results}), we compared different strategies of overfitting the OMPs. For simplicity, we trained on a subset of the CLIC training set, and evaluated on the CLIC validation set. In one strategy ($c_1$), that we adopted in our final codec, 
we overfit a few selected layers of last DSA block in the intra-frame decoder 
by using the first intra frame of each video, and then employ the OMPs for all intra frames of that video; 
in another strategy ($c_2$), we overfit the OMPs of all layers of last DSA block, and the very last layer of the decoder, 
separately on each intra frame. The Score was computed by using $\lambda=0.15$. 
In the strategy $c_2$, the overfitted OMPs still required a much higher bitrate overhead after quantization. 
Since we include the quantized OMPs into bitstream only 
if the parameters provide coding gains in terms of the overall loss (over a video for $c_1$ and over the considered frame for $c_2$), 
the strategy $c_1$ provided smaller bitrate overhead and better MS-SSIM than $c_2$ on average.
Compared to the size of latent tensor bitstreams, the size of quantized-OMPs bitstreams is only approximately $0.06\%$ for the CLIC test set 
and $0.008\%$ for the JVET-CTC sequences. 
Moreover, we evaluated the performance of the intra-frame codec without 
overfitting the OMPs ($c_3$ in Table \ref{tab:study_results}), which confirms the benefits of our proposed technique. 

\begin{table}
\caption{Additional studies.}
\begin{center}
\begin{tabular}{|c|c|c|c|c|c|}
\hline
\multicolumn{6}{|c|}{\textbf{Intra codec only}} \\
\hline
\textbf{ID} & \textbf{Overfit} & \textbf{Overfit} & \textbf{BPP} & \textbf{MS-SSIM} & \textbf{Score}\\
& \textbf{layers} & \textbf{frames} & \multicolumn{3}{|c|}{}\\
 \hline
$c_1$ & \textit{Few} & \textit{First} & $0.11190$ & $0.98550$ & $0.96872$ \\
$c_2$ & \textit{All} & \textit{All} & $0.11196$ & $0.98544$ & $0.96864$ \\
$c_3$ & \textit{None} & \textit{None} & $0.11175$ & $0.98539$ & $0.96863$ \\
\hline\hline
\multicolumn{6}{|c|}{\textbf{Inter codec only}} \\
\hline
\textbf{ID} & \textbf{Overfit} & \textbf{Scaling} & \textbf{BPP} & \textbf{MS-SSIM} & \textbf{Score}\\
 & \textbf{Combiner} & \textbf{motion} & \multicolumn{3}{|c|}{}\\
\hline
$c_4$ & - & - & $0.01201$ & $0.98487$ & $0.98367$ \\
$c_5$ & - & \checkmark & $0.01913$ & $0.99179$ & $0.98988$ \\
$c_6$ & \checkmark & - & $0.01216$ & $0.98589$ & $0.98468$ \\
\hline

\end{tabular}
\label{tab:study_results}
\end{center}
\end{table}

We also performed another ablation study involving the inter frame codec (see \textbf{Inter codec only} part in Table \ref{tab:study_results}), 
which was trained on a subset of the CLIC training set, and evaluated on 5 separate videos.
We evaluated the contribution of overfitting the scaling parameters and of resolution adaptation. 
$c_4$ is our inter-frame codec without overfitting the Combiner's scaling parameters and without resolution-adaptation scaling of motion;
$c_5$ is our inter-frame codec without overfitting the Combiner's scaling parameters but with resolution-adaptive scaling of motion; 
$c_6$ is our inter-frame codec with overfitting the Combiner's scaling parameters but without resolution-adaptive scaling of motion.
$\lambda=0.1$ was used to measure the Score. The ablation study results demonstrate that the proposed techniques clearly improve the coding performance.

\section{Conclusions}
In this paper, we proposed a set of techniques for enabling a learned video codec to be highly adaptive to the input content, thus overcoming potential limitations caused by domain shift. Via extensive experiments, we showed the benefits of our techniques, and compared our codec to state-of-the-art video codecs in different settings and for different datasets. 


\bibliographystyle{IEEEtran}
\bibliography{egbib}

\end{document}